\documentclass[aps,prd,aapm,mph,amsmath,amssymb,reprint]{revtex4-2}

\usepackage{graphicx}  
\usepackage{bm}        
\usepackage{amssymb}   
\usepackage{amsmath}   
\usepackage{color} 
\usepackage{lineno}
\usepackage{multirow}
\usepackage{enumerate}
\usepackage{comment}
\usepackage{tabularx}
\usepackage{booktabs}
\usepackage{natbib}
\usepackage{array}
\usepackage{soul}

\newcolumntype{P}[1]{>{\centering\arraybackslash}p{#1}}
\newcolumntype{Y}{>{\centering\arraybackslash}X}

\usepackage{gensymb}

\newcommand{\dd}{{\rm d}}

\def\lapp{\ifmmode\stackrel{<}{_{\sim}}\else$\stackrel{<}{_{\sim}}$\fi}
\def\gapp{\ifmmode\stackrel{>}{_{\sim}}\else$\stackrel{<}{_{\sim}}$\fi}
\setcitestyle{authoryear, open={(},close={)}, yysep={,}}

\usepackage{dcolumn}
\usepackage{bm}

\begin{document}

\title{Study on the detectability of gravitational radiation from single-binary encounters between black holes in nuclear star cluster: the case of hyperbolic flybys.}

\author{Elena Codazzo}
\email{elena.codazzo@gssi.it}
\affiliation{Gran Sasso Science Institute (GSSI), I-67100 L’Aquila, Italy}
\affiliation{INFN, Laboratori Nazionali del Gran Sasso, I-67100 Assergi, Italy}
\author{Matteo Di Giovanni}
\affiliation{Gran Sasso Science Institute (GSSI), I-67100 L’Aquila, Italy}
\affiliation{INFN, Laboratori Nazionali del Gran Sasso, I-67100 Assergi, Italy}
\author{Jan Harms}
\affiliation{Gran Sasso Science Institute (GSSI), I-67100 L’Aquila, Italy}
\affiliation{INFN, Laboratori Nazionali del Gran Sasso, I-67100 Assergi, Italy}
\author{Marco Dall'Amico}
\affiliation{Universit\`a di Padova, Dipartimento di Fisica e Astronomia, I-35131 Padova, Italy}
\affiliation{INFN, Sezione di Padova, I-35131 Padova, Italy}
\author{Michela Mapelli}
\affiliation{Universit\`a di Padova, Dipartimento di Fisica e Astronomia, I-35131 Padova, Italy}
\affiliation{INFN, Sezione di Padova, I-35131 Padova, Italy}

\begin{abstract}
With the release of the third Gravitational-Wave Transient Catalogue (GWTC-3), 90 observations of compact-binary mergers by Virgo and LIGO detectors are confirmed. Some of these mergers are suspected to have occurred in star clusters.
The density of black holes at the cores of these clusters is so high that mergers can occur through a few generations forming increasingly massive black holes. These conditions also make it possible for three black holes to interact, most likely via  single--binary encounters. In this paper, we present a first study of how often such encounters can happen in nuclear star clusters (NSCs) as a function of redshift, and whether these encounters are observable by gravitational-wave (GW) detectors. This study focuses on effectively hyperbolic encounters leaving out the resonant encounters. We find that in NSCs single--binary encounters occur rarely compared to binary mergers, and that hyperbolic encounters most likely produce the strongest GW emission below the observation band of terrestrial GW detectors. While several of them can be expected to occur per year with peak energy in the LISA band, their amplitude is low, and detection by LISA seems improbable. \\
\noindent {\bf Keywords: gravitational waves, single-binary encounters, hyperbolic encounters, star clusters}
\end{abstract}

\maketitle

\section{Introduction}
The constant upgrades to current-generation gravitational-wave detectors Advanced Virgo \citep{aVirgo} 
and Advanced LIGO \citep{aLIGO} and their much improved sensitivities made the detection of GW events a common occurrence. During the last few years, the observation of several tens of GW signals from compact object mergers \citep{gwtc1, gwtc2, gwtc3} has started a new era in GW astronomy. The first observation of a binary neutron star merger also proved the feasibility of joint GW and electromagnetic observations, thus opening the era of multimessenger astronomy \citep{bns, multim}.

With the successful outcome of current-generation detectors, the GW community is planning the construction of next-generation detectors, such as Einstein Telescope (ET) \citep{et,ET2020} and Cosmic Explorer \citep{EvEA2021}. The improved sensitivity of these ground-based detectors is expected to increase the number of observed events from tens per year to hundreds of thousands per year opening an enormous science case \citep{MaEA2020,KaEA2021a}. Moreover, the addition of the planned space-based detector LISA \citep{ASEA2017} together with pulsar-timing arrays \citep{HoEA2010} and possibly with decihertz, Moon-based GW detectors \citep{gloc, lgwa} will enable the scientific community to cover the GW spectrum from nanohertz to kilohertz.

Both the positive results of current GW astronomy and the perspective of more sensitive future detectors, which will broaden the accessible frequency band of the GW spectrum, prompted a series of studies aimed at establishing the possibility of detecting GWs from non-canonical sources (i.e., different from compact object mergers, continuous waves and a stochastic background). Among these sources are hyperbolic encounters between compact objects, black holes (BHs) in particular. This kind of interactions on unbound orbits between isolated BHs have already been the subject of studies with the scope of providing both the analytical \citep{8capozziello2008gravitational,majar2010gravitational,9de2012gravitational,de2014gravitational,cho2018gravitational,99garcia2018gravitational,mitra2021detectability,morras2022search} and numerical \citep{damour2014strong,nagar} tools to determine the GW emission of a fly-by or of a dynamical capture.
Dynamical friction in star clusters (SCs) causes the segregation of BHs, and the very high densities reached in the SC core triggers both the formation of binary black holes \citep[BBHs; see, e.g.,][for a review]{Mapelli2020} and close encounters with other stars and BHs \citep{portegies2000,banerjee2010,tanikawa2013,oleary, mapellietal2013, ziosietal,rodriguez2015,rodriguez2016,rodriguez2019, mapelli2016,askar2017, seddaetal,samsing2018,samsing2018b,fragione2018,fragione2019,fragione2022,zevin2019,zevin2021,kremer2019, Sedda_2020, mapellietal2021, rastello,banerjee2021,rizzuto2022,kamlah2022}.  
 Therefore, since the rate of binary--single encounters scales with the local density of stars \citep{sigurdsson1993},   
we also expect a significant contribution from BBH-BH triple encounters in dense stellar environments. However, the complexity of extending the formalism of hyperbolic encounters between BHs to binary--single encounters results in a significant lack of studies aimed at characterizing BBH-BH events. Therefore, in this paper we propose 
a numerical method to estimate the GW emission and the rate of binary--single hyperbolic encounters in nuclear star clusters. The questions to be answered are whether these signals are likely to be detected with current or future GW detectors either as individual signals or as a stochastic background.

This paper is organized in the following way. In Section \ref{section2}, we give an overview of the underlying astrophysical assumptions of the systems considered in this study; in Section \ref{section3} we discuss in detail the simulations done for this work and the method used to study BBH-BH encounters. The results are presented and discussed in Section \ref{section4}. A short summary concludes the paper in Section \ref{section6}.

\section{Astrophysical background}\label{section2}
Initially, following the work by \cite{spitzer1967}, globular clusters (GC) were not thought to retain a significant number of BHs because, due to the much higher mass of BHs compared to typical stars, the BHs would quickly mass segregate to form an isolated subcluster that is dynamically decoupled from the GC. Due to the small size, high density, and small number of objects in the subclusters, relaxation and strong encounters were expected to eject the majority of BHs on a timescale of 1\,Gyr.

Nevertheless, recent works 
 have changed the picture and now predict that large numbers of BHs can remain bound to star clusters where they interact to form binaries \citep[e.g.,][]{ziosietal,breen2013, morscher2015,rodriguez2015,mapelli2016,askaretal2019, dicarloetal, rastello}. In particular, for our case study, we choose to focus on NSCs where mass segregation and the high density reached in their cores favours the formation of BBHs and three-body encounters \citep{Mapelli2020}.

\subsection{Properties of nuclear star clusters}

NSCs are the oldest among the different types of clusters, with an estimated age of $\sim$13.6\,Gyr \citep{neumayer}, and are defined as extremely dense and massive star clusters occupying the innermost region or nucleus of most galaxies. From an observational point of view, NSCs are identified as luminous and compact sources that clearly stand out above their surroundings \citep{neumayer}. It has also been argued that, for lower masses, NSCs are formed primarily from GCs that inspiral into the center of the galaxy \citep{tremaine1975formation,capuzzo1993evolution,antonini2012dissipationless,pfeffer2018mosaics}, while for higher masses star formation within the nucleus forms the bulk of the NSC \citep{hopkins2010nuclear,hopkins2010massive,mapelli2012situ, guillard2016new}.

The two-body relaxation time-scale of NSCs, i.e., the time that a cluster needs to reach thermal equilibrium through two-body encounters, is related to the half-mass radius $r_{\rm h}$ of the cluster and is defined as \citep{spitzer97}:

\begin{equation}
t_{\rm rh} \approx 4.2\times10^9 \left(\frac{15}{\ln\Lambda} \right) \left(\frac{r_{\rm h}}{4\,\rm pc} \right)^{3/2} \left( \frac{M_{\rm cl}}{10^7M_{\odot}} \right) ^{1/2}\,\rm yr,
\end{equation}

with $\ln{\Lambda} \sim$10 being the Coulomb logarithm and M$ _{\rm cl} $ the total mass of the cluster. Even though for the most massive NSCs this time may be higher than the Hubble time, BHs still manage to segregate in the core on a lower timescale, defined as the dynamical friction timescale \citep{Chandrasekhar43}:

\begin{equation}
t_{\rm df} \approx \dfrac{3}{4(2\pi)^{1/2} G^2 \ln\Lambda} \dfrac{\sigma^3}{\rho\, m_{\rm BH}},
\end{equation}

where $\sigma$ is the 3D velocity dispersion, $m_{\rm BH}$ the mass of the BH and $ \rho=3M_{\rm cl}/(8\pi$r$^3_{\rm h})$ the mass density at the half-mass radius. 
In this way, the core reaches very high densities, of the order of $\sim10^6$\,pc$^{-3}$, favoring close encounters between its components.

The assumption we make is that our BHs dynamically evolve in a cluster whose properties are stationary.

\subsection{Properties of binary black holes in star clusters}

Binary systems are hard or soft according to their binding energy E$_{\rm b}$ \citep{heggie1975binary}. Soft binaries have a binding energy less than the average kinetic energy of the stars in the cluster; hard binaries have a binding energy higher than the average kinetic energy of the stars in the cluster. All the binaries we refer to in this work are hard binaries, i.e.:
\begin{equation}
E_{\rm b}= \dfrac{G m_1 m_2}{2a} \gtrsim \frac{1}{2} m_{\star} \sigma^2,
\end{equation}
where $m_1$ and $m_2$ are the primary and the secondary mass of the BBH, $a$ is the semi-major axis of the binary, $m_{\star}$ is the mass of a typical star in the cluster and $\sigma$ is the three-dimensional dispersion velocity.

The dynamical formation of BBHs in clusters is possible through different formation channels.  
The fastest way through which BBHs can form is via three-body encounters between three isolated BHs \citep{Mapelli2020, Sedda_2020, fragione2020repeated}. These encounters happen in high-density conditions, such as during the core-contraction of the cluster.
A temporary triple system is formed, which will result in a binary plus an ejected object. The timescale of this process is \citep{lee1995evolution}:
\begin{equation}\begin{split}
t_{\rm 3bb} \approx 125\,\text{Myr} \left( \dfrac{10^6\, \rm pc^{-3}}{n_{\rm c}} \right)^2 \left( \zeta^{-1} \dfrac{\sigma_{\rm 1d}}{30\,\rm km\,s^{-1}} \right)^9 & \\  
\cdot\left(\frac{20M_{\odot}}{m_{\bullet}} \right)^5& , 
\end{split}\end{equation}
where $n_{\rm c}$ is the central density of the cluster; $\sigma_{\rm 1d}=\sigma/\sqrt{3}$ is the one dimensional velocity dispersion at $r_{\rm h}$, assuming that the stellar velocities are isotropically distributed; $\zeta \leq 1$ is a constant that takes into account the deviation from the equipartition of the system; $\zeta= 1$ means that there is equipartition, and this is the case we consider; $m_{\bullet}$ is the mass of a massive BH with velocity dispersion $\sigma_{\rm BH}$ according with the relation $\zeta m_{\bullet} \sigma_{\rm BH}^2=m_{\star} \sigma^2$.

Another possible scenario is the formation of a new BBH through BH exchange in an original binary through a binary--single encounter. 
These encounters are more likely to happen when the fraction of binaries in the cluster is high. Replacement of a BH in such an encounter makes the system energetically more stable. This formation mechanism is slower than the previous one, with its timescale being \citep{miller2009mergers}:
\begin{equation}\begin{split}
t_{\rm sb} \approx 3\,\text{Gyr} \left( \frac{0.01}{f_{\rm bin}}\right) \left( \dfrac{10^6\,\rm pc^{-3}}{n_{\rm c}} \right) \left(\frac{\sigma}{30\,\rm km\,s^{-1}}  \right)&\\ 
\cdot\left( \dfrac{10M_{\odot}}{m_{\rm tot}}  \right)  \left(\dfrac{1\,\rm AU}{a_{\rm hard}}  \right)& , 
\end{split}\end{equation}
where f$_{\rm bin}$ is the binary fraction; $m_{\rm tot}$ is the sum of the three masses of the system; $a_{\rm hard}$ is the typical semimajor axis of a hard binary.

In the densest clusters, there is also a third possible formation channel for BBHs, i.e., through two-body captures, where two isolated BHs interact to form a binary \citep{quinlan1990dynamical}. The timescale for this phenomenon is \citep{quinlan1990dynamical}: 

\begin{equation}\begin{split}
 t_{\rm cap} \approx 7.7\times{} 10^3\,\text{Gyr} \left( \dfrac{M_{\odot}}{m_{\bullet}} \right)^2  \left( \dfrac{10^8\,\rm pc^{-3}}{n_{\rm c}} \right)& \\ \cdot\left(\frac{\sigma}{200\,\rm km\,s^{-1}}. \right)^{11/7}&
\end{split} \end{equation} 

For our study, we assume that, at the beginning of the simulations, the formation process of BBHs in the clusters is over, regardless of the formation channel. Therefore we neglect any contribution from the encounters prior to the formation of BBHs.

Once BBHs are formed in the clusters, they harden at a constant rate through binary--single encounters \citep{heggie1975binary}.
In general, flybys are the majority outcome of such interactions and since, statistically, the velocity of the intruder after the encounter is greater than the one with which it approached the binary, as a consequence of the conservation of energy, the binary tightens according to Heggie's law that states that hard binaries tend to become harder.

The semi-major axis of the binary will decrease over time due to binary--single encounters as follows:
\begin{equation}\label{dynamical_hardening}
\dfrac{\text{d}a}{\text{d}t}\big|_{\rm 3b}= -2 \pi \xi\dfrac{G \rho_{\rm c}}{\sigma} a^2,
\end{equation}
where $ \rho_{\rm c} $ is the local density of stars and $\xi\approx $ 3 is a dimensionless hardening rate \citep{quinlan1996dynamical}.
This contribution of hardening proportional to $a^2$ dominates over that due to the GW emission by the system, which instead takes over when the semi-major axis is small since it is proportional to $a^{-3} $ \citep{peters1964gravitational}:
\begin{equation}
\dfrac{\text{d}a}{\text{d}t}\big|_{\rm GW}=-\frac{64}{5} \dfrac{G^3 m_1 m_2(m_1m_2)}{c^5a^3(1-e^2)^{7/2}}f(e),
\end{equation}
with
\begin{equation}
f(e)=\left( 1+\frac{73}{24}e^2 + \frac{37}{96}e^4 \right).
\end{equation}
The evolution of the semi-major axis can be written as the sum of the two contributions \citep{Mapelli2020} 
\begin{equation}\label{3b+gw}
\dfrac{\text{d}a}{\text{d}t}=\dfrac{\text{d}a}{\text{d}t}\big|_{\rm 3b}+\dfrac{\text{d}a}{\text{d}t}\big|_{\rm GW}=-c_1 a^2 - \frac{c_2}{a^3},
\end{equation}
where
\begin{equation}
c_1=2 \pi \xi\dfrac{G \rho_c}{\sigma}, \,\,\,\,\, c_2=\frac{64}{5} \dfrac{G^3 m_1 m_2(m_1m_2)}{c^5(1-e^2)^{7/2}}f(e).
\end{equation}
The value of $a$ at which the transition between the two regimes occurs is obtained by imposing $(\dd a/\dd t)_{\rm 3b}= (\dd a/\dd t)_{\rm GW} $:
\begin{equation}
a_{\rm GW}=\left[ \dfrac{32G^2}{5\pi\xi c^5} \,{}\dfrac{\sigma m_1m_2(m_1+m_2)}{\rho_c (1-e^2)^{7/2}} f(e) \right] ^{1/5}.
\end{equation}



\section{Method and simulations}\label{section3}
We obtain the rates of BBH-BH encounters and an estimate of the emitted GW spectrum with numerical simulations.  We use the N-body  simulation code ARWV \citep{mikkola1993implementation, chassonnery2, chassonnery1} to simulate the three-body encounters to characterize the possible types of encounters and as benchmark for simplified analytical models. The number of encounters happening in NSCs is estimated by use of a Monte-Carlo simulation starting from appropriate initial conditions.

\subsection{Initial conditions}\label{initial_conditions}
The initial conditions for single--binary encounters are characterized by the parameters and formalism first introduced by \cite{hutbahcall} and then updated by \cite{dall2021gw190521}. The orientation angles of the encounter are drawn randomly from an isotropic sphere and are defined as: 
\begin{itemize}
    \item $\phi \in [0,2\pi)$ is the angle between the pericenter of the binary orbit and the intersection of the vertical plane in which lies the initial velocity of the intruder;
    \item $\psi\in [0,2\pi)$ is the orientation of the impact parameter with respect to the orbital plane direction in a surface perpendicular to the initial velocity of the intruder;
    \item $\theta : \cos (\theta) \in [-1;1]$
    is the angle between the direction perpendicular to the orbital plane and the intruder initial velocity at infinity.
\end{itemize}
The orbital phase is generated in the range $[-\pi;\pi]$.

We sampled the initial magnitude velocity of the intruder from a Maxwell-Boltzmann distribution with a dispersion velocity of 50\,km/s, which is typical of NSCs \citep{georgievetal2016}. We generate the binary orbital eccentricities based on a thermal distribution proportional to $e^2$ in the range [0,1] \citep{heggie1975binary}.

The initial distance between the center-of-mass of the binary and the intruder is set to be 100 times the initial semi-major axis of the binary so that the intruder does not feel the gravitational potential of the binary at the beginning of the simulation. The impact parameters $b$ are sampled from a distribution proportional to $b^2$ with limits $[0,b_{\rm max}]$ where $b_{\rm max}$ is derived from \cite{sigurdssonIMPACT}:
\begin{equation}
    b_{\rm max} = \frac{\sqrt{2\,{}G\,{}(m_1 + m_2 + m_3)\,{}a}}{v_{\infty}}.
\end{equation}
Furthermore, the values of $b$ are kept only if $b$ is smaller than the initial distance $D$ of the intruder black hole $m_3$; the opposite situation is geometrically unrealistic. 

For what concerns the parameters of the binary, the semi-major axis is generated with a uniform distribution from 0 to 1000\,AU and then rejected if the binary is soft or if the binary is too hard. If it is too hard (small), it would merge in the first timesteps of the ARWV simulation or it would be ejected from the cluster as a consequence of an extremely hard interaction. The initial conditions for the three masses are drawn from the distributions presented by \cite{dicarloetal} based on the astrophysical evolution of stars in a dense environment, assuming the metallicity of the cluster to be 0.002. 
We sampled the spins of the three BHs from a Maxwellian distribution with root mean square 0.1.

\subsection{Evolution of binary parameters}\label{evo}
After defining the parameters that characterize an encounter and a binary system, we simulate the evolution of BBHs in NSCs to assess how many of them are retained in a cluster, after a given amount of time, and therefore provide the number of encounters that we have to simulate. The procedure consists in considering an initial set of binaries and to let them evolve following equation (\ref{3b+gw}), within the time window $[0,t_{\rm NSC}]$
, where $t_{\rm NSC}$ is the current life time of the cluster and 0 is the time at which a BBH is already formed, i.e., after a time $t=t_{\rm df}+\min(t_{\rm 3bb}, t_{\rm tsb}, t_{\rm cap})$. We find that the time window between the first formation of an hard BBH inside a NSC and t$_{\rm NSC}$ is about 12\,Gyr; therefore each simulation, if not interrupted, will last 12\,Gyr. 


The semi-major axis of a binary, according to equation (\ref{dynamical_hardening}), evolves due to binary--single encounters and decreases at a constant rate \citep{heggie1975binary}. We consider that, after each encounter, the binding energy of the binary increases by a quantity:
\begin{equation}
\Delta E_b = \xi \dfrac{m_3}{m_1+m_2} E_b,
\end{equation} 
thus varying the semi-major axis according to
\begin{equation}
a_{\rm new} = -G \dfrac{m_1m_2}{2E_{\rm new}}.
\end{equation} 
This helps to keep track of how often an encounter occurs according to equation (\ref{3b+gw}). For each time step, the simulation guarantees that the new time does not exceed $t_{\rm NSC}$ and that the binary does not have enough velocity to escape from the cluster. This last check is done on the semi-major axis: the minimum semi-major axis that a binary can have without being ejected from the cluster due to dynamics is \citep{coleman2002production}
\begin{equation}
a_{\rm ej}=\dfrac{2\xi m_\star^2}{(m_1+m_2)^3} \dfrac{G m_1m_2}{v^2_{\rm esc}}.
\end{equation}
When the semi-major axis decreases to $a_{\rm ej} $, we compare it to $a_{\rm gw}$ i.e. its value at which the emission of gravitational waves from the binary become dominant. If $a_{\rm ej} < a_{\rm gw}$, the binary will merge into the cluster before it can be ejected; otherwise the binary will be ejected from the cluster, and we will no longer take it into account in the simulation. We considered a delay time of 1\,Gyr after the merger, according to Sec \ref{section2}, required for the resulting BH to drift back into the core and form a new hard binary: when a binary in the cluster merges, the resulting BH experiences a relativistic kick that ejects it outside the star cluster's core. After each merger, however, we check if $v_{\rm kick}<v_{\rm esc}$ before proceeding with the integration, where $v_{\rm kick}$ is the relativistic kick \citep{lousto2012gravitational} and $v_{\rm esc}$ is the escape velocity from the NSC.

The kick depends on the mass ratio and spins of the progenitor BHs. To compute it we draw the spins of the two BHs according to a Maxwell distribution with one-dimensional root-mean square 0.1, as inferred from GWTC-2 \citep{abbott2021population} and with random direction since they are in a dynamical environment. We then calculated the kick following \cite{lousto2012gravitational}:
\begin{equation}
v_{\rm kick}=(v_{\rm m}^2 + v^2_{\perp} +2v_{\rm m} v_{\perp} \cos\phi +v^2_{\parallel})^{1/2},
\end{equation}
where
\begin{equation}
\begin{aligned}
v_{\rm m}=&A \eta^2 \dfrac{1-q}{1+q} (1+B\eta) \\
v_{\perp}=&H \dfrac{\eta^2}{1+q} \big|\chi_{1\parallel} -q \chi_{2\parallel}\big| \\
v_{\parallel}=&\dfrac{16\eta^2}{1+q} \left[ V_{1,1}+V_{\rm A} S_{\parallel} +V_{\rm B} S_{\parallel}^2+V_{\rm C} S_{\parallel}^3 \right]\\ &\big|\chi_{1\perp} -q \chi_{2\perp}\big| \cos(\phi_\Delta-\phi)
\end{aligned}
\end{equation}
with $q = m_2/m_1$ assuming $m_2 \leqslant m_1$, $\eta = q(1+q)^{-2}$, $A=1.2\times 10^4$\,km\,s$^{-1}$, $B=-0.93$, $H=6.9\times10^3$\,km\,s$^{-1}$ and $V_{1,1}=3678$\,km\,s$^{-1}$, $V_{\rm A}=2481$\,km\,s$^{-1}$, $V_{\rm B}=1792$\,km\, s$^{-1}$ and $V_{\rm C}=1506$\,km\,s$^{-1}$. Vectors with subscripts $ \parallel $ and $ \perp $ are respectively parallel and perpendicular to the orbital angular momentum. $\chi_1$ and $\chi_2$ are the spin vectors relative to the two black holes, and the vector $\vec S$ is defined as $\vec{S} = 2(\vec{\chi}_1+q^2\vec{\chi}_2)/(1+q)^2$. The angle $\phi$ is the phase of the BBH that we have randomly taken between 0 and 2$\pi$, and $\phi_\Delta$ is the angle between the in-plane component of the vector $\vec{\Delta} \equiv (m_1+m_2)^2(\vec{\chi}_1+q\vec{\chi}_2)/(1+q)$ and the infall direction at merger.

If the resulting BH is retained in the cluster, we consider its total mass as 95\% of the mass of the two progenitor BHs, in order to take into account the emitted gravitational radiation, and we randomly take its spin in the interval [0.6,0.9]. We form a new generation binary with this BH and a BH coming from our distribution of singles BH with mass higher than 10\,M$_\odot$, to prevent lighter BHs from being replaced in the first few encounters of the binary \citep{heggie1975binary}. The semi-major axis is uniformly generated between the values $a_{hard}$ at which the binary start to be hard and its $a_{crit}$ i.e., the max($a_{gw}$, \,$a_{ej}$).

\subsection{Number of simulated binaries}\label{N.E.}

\begin{figure}[t]
\includegraphics[width=\columnwidth]{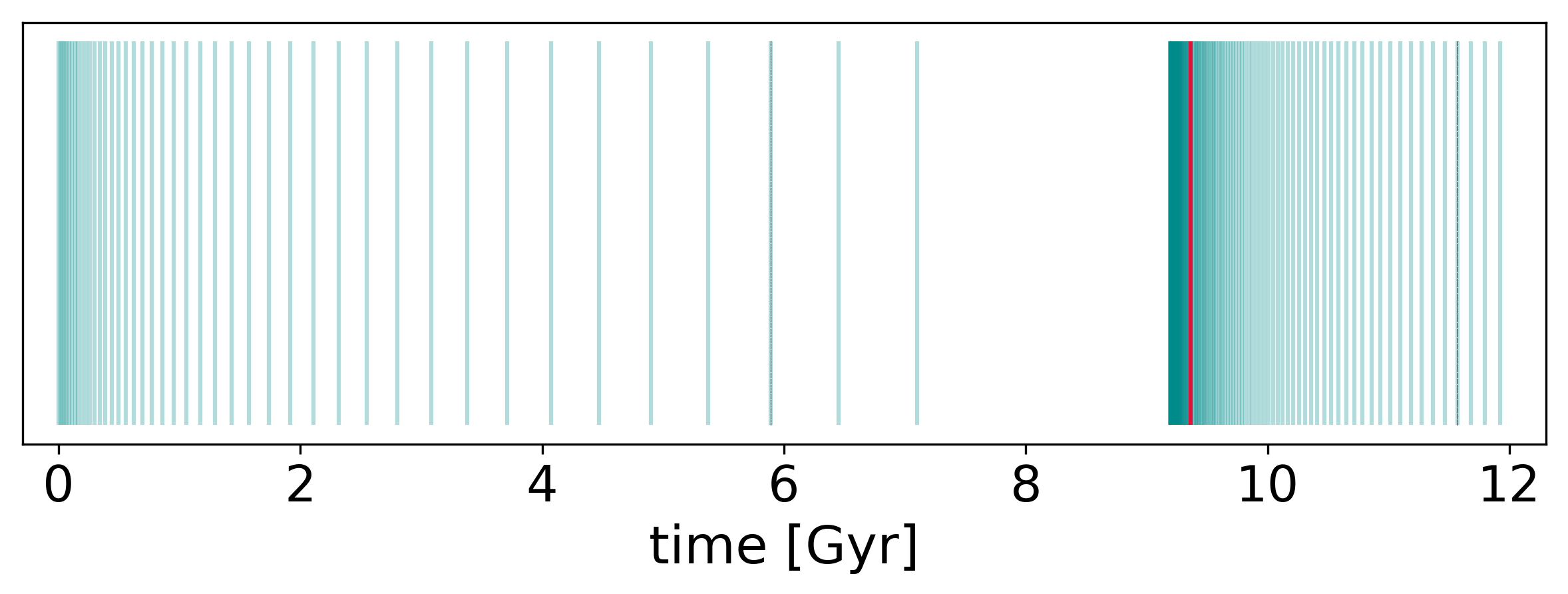}
\caption{Evolution of a BBH in 12\,Gyr for the scenario in which the core of the NSC has 1\,pc radius. The green lines refer to an encounter with a star, while the red line to an encounter with a BH. 
There are two generations. In the second generation, starting around 9\,Gyr, one component of the binary is the BH
resulting from the merger of the binary of first generation that after 1 Gyr of time delay has formed a new hard binary.
The dashed gray lines refer to the time when the semi-major
axis of that binary is a$_{\rm gw}$. The simulation ends at current
time (12\,Gyr).}
\label{encounters}
\centering
\end{figure}

Rough estimates of the number of binaries present at the formation of each NSC can be obtained by considering an average mass of the cluster of $M_{\rm cl} = 1.5\times{}10^6 M_\odot$ with a total mass in BHs of $M_{\rm bh}=0.001 \,{} M_{\rm cl}$. For a typical mass of the BHs of $10\, M_{\odot}$, the number of BHs in each NSC is about $N _{\rm BH} =1500$. We consider a fraction of $0.01\,{} N_{\rm BH}$ of BBHs \citep{antoninirasio} and an average density of galaxies of 0.03\,Mpc$^{-3}$ from 60\,Mpc up to redshift 3.5 and, for distances less than 60\,Mpc, we follow Eq. 6 of \cite{kocsis2006detection}. 
Following the procedure described in Section \ref{evo}, we simulate our sets of initial binaries in two different scenarios: one in which the size of the core of the NSC has a radius of 0.1\,pc and one in which it is 1\,pc. In both cases, our population of BHs dwells in the core, where we have assumed a constant density of objects of $10^6$~pc$^{-3}$.

In Figure \ref{encounters}, there is an example of a BBH evolution over 12\,Gyr assuming the core of the NSC to have a radius of 1\,pc. Each encounter is represented by either a green line if the intruder is a star or a red line if the intruder is a BH; we are interested in encounters between the BBH with other BHs. At the time $t=0$ a binary from the initial set of binaries starts to evolve, decreasing its semi-major axis according to (\ref{3b+gw}). The dashed gray lines refers to the moment when the semi-major axis reaches the value max($a_{\rm gw}$, $a_{\rm ej}$). In the specific case of Figure \ref{encounters}, the binary merges around 7.5\,Gyr. The resulting black hole is retained in the cluster and manages to find a new BH companion, forming a second-generation binary. The latter also manages to evolve in the cluster making some encounters up to the current times. Between the merger and the formation of the second generation binary, there is a delay time of 1\,Gyr.

\subsection{Simulating 3-body encounters with ARWV}\label{ARW}
\begin{figure}[t]
\includegraphics[width=0.98\columnwidth]{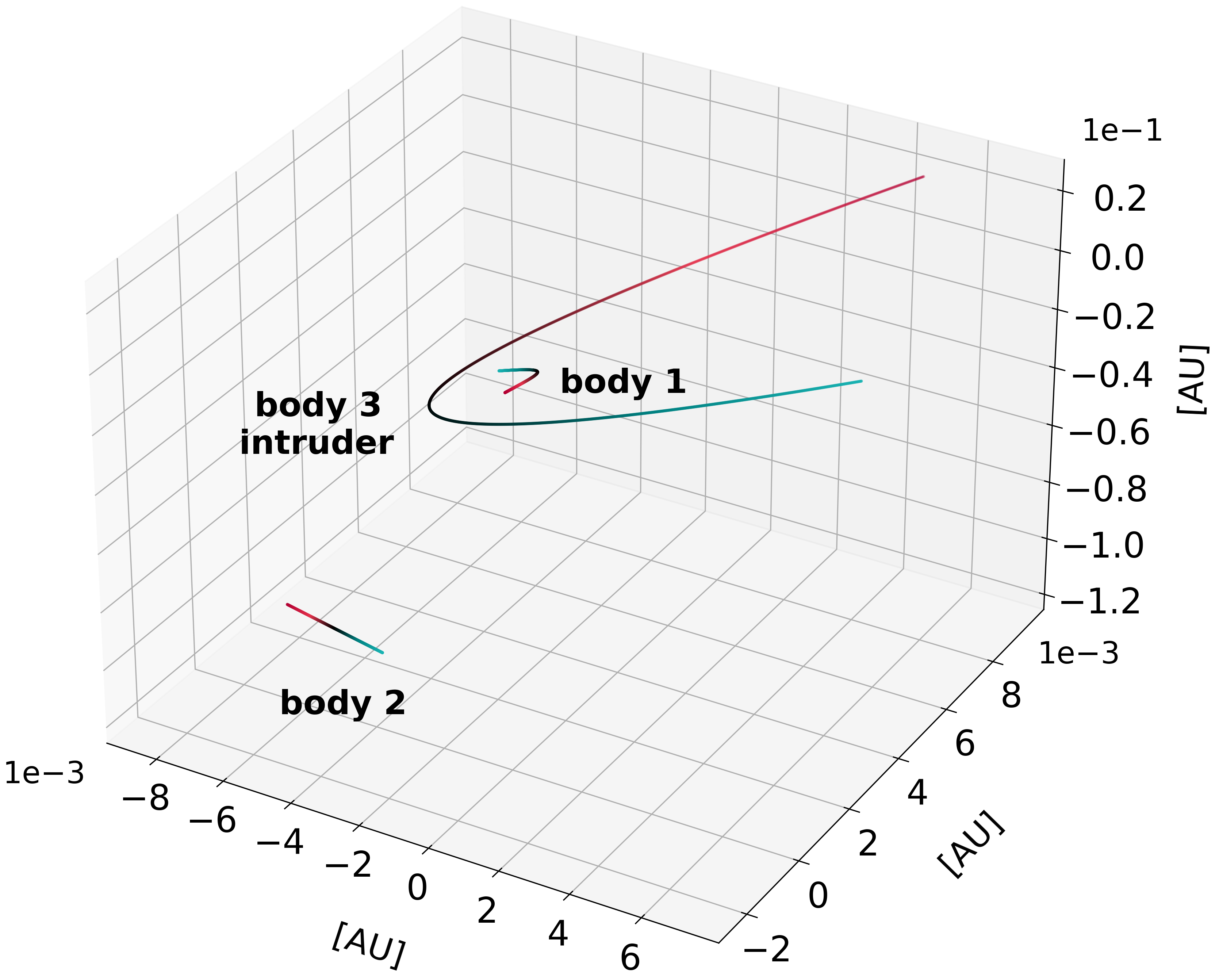}
\caption{Trajectory of a single--binary encounter. The encounter in the plot lasts 16 minutes and has a time resolution of 1s, starts when the trajectory is blue and ends when it is red. Body 1 and body 2 are the components of the binary. The masses for body 1, body 2 and body 3 are 94.5\,M$_{\odot}$, 10.9\,M$_{\odot}$ and 7.1\,M$_{\odot}$, respectively. The closest approach between body 1 and body 3 is 0.002\,AU with a relative velocity between the two of $8\times 10^3$\,km/s; at that moment the binary separation is 0.128\,AU.}
\label{trajectory}
\centering
\end{figure}

The dynamics of each BBH-BH encounters are simulated, individually, through multiple runs of the direct N-body code ARWV (\citealt{chassonnery2}; \citealt{chassonnery1}). ARWV makes use of the algorithmic regularization chain method to integrate the equations of motion (\citealt{mikkolaEOM1989}; \citeyear{mikkolaEOM1993}). The use of this chain scheme reduces the round-off errors making the regularization algorithm more efficient, especially for close interactions. ARWV also includes a post-Newtonian treatment up to order 2.5 to correct the equations of motion in case of strong gravitational interactions \citep{mikkolaPN2008}.

Equation \ref{3b+gw} tells us when an encounter with a BH takes place during the simulation of the evolution of BBHs in the NSC. This determines the redshift at which the encounter happens, the semi-major axis of the binary at the moment of the encounter, and the masses of the three bodies. We use this information as the initial conditions for ARWV simulations. The parameters of the encounter (spins of the three BHs, velocities, impact parameters and the angles $\phi, \psi$ and $\theta$) are drawn randomly from their distributions described in Section \ref{initial_conditions}. ARWV outputs the positions and velocities of each body at each time step. In this way, it helps us to discriminate whether the initial parameters of the BBH-BH system lead to an encounter that is a simple flyby or a resonant encounter - multiple flybys around the binary -, in which the intruder can also take the place of one of the two components of the binary for some orbits or permanently. 


For the present work we will consider only flybys, i.e., those encounters in which the intruder follows a nearly hyperbolic trajectory around the binary or, more frequently, across it. Resonant encounters will be included in the study at a later stage, because of their intrinsic complexity. Indeed, they happen on a much longer timescale: from our ARWV simulations we observed that, on average, the shortest duration of a resonant encounter is in the range of days, making it much more computationally expensive to simulate them at a fine-grained temporal resolution. Moreover, as we will detail in Sec.~\ref{signal}, when we compute the gravitational spectrum we obtain noisy results at high frequencies. Signals resulting from hyperbolic encounter tend to be similar in shape, while there could be more diverse waveforms coming from resonant encounters given the wide variety of orbits they could describe. Therefore, we will need to adopt a different approach with respect to the one described in Sec.~\ref{signal}; this will be the subject of future developements of this work.


From the output of the ARWV simulations with a time resolution of 1\,s we identified the moment of maximum approach between the intruder and one of the two components of the binary, since in almost all cases, at the time of the encounter, the distance between the two components of the binary is at least double the distance between the intruder and the object around which it flies by. This allowed us to roughly identify the characteristic frequency of the event as
\begin{equation}
f=\frac{1}{2\pi} \frac{v'}{r_{\rm min}} \label{f},
\end{equation}
neglecting the presence of the third body. Here, $r_{\rm min}$ is the minimum distance between the intruder and the closest body of the binary while $v'$ is the relative velocity when the minimum distance is reached. We use this frequency value to adjust the time resolution and repeat the simulation until a satisfactory representation of the trajectories is obtained around the closest encounter, like in the case of Figure \ref{trajectory}.

\section{Results}\label{section4}

In our calculations, the time evolution of the semi-major axis of the binaries in NSCs is driven by two processes: encounters with other bodies in the surrounding environment and emission of gravitational waves. From these simulations, when an encounter with another black hole occurs, we extract the main parameters that describe a single--binary encounter between black holes: the masses of the binary, its semi-major axis, the mass of the intruder and the time at which the encounter happens in the simulation, from which we then obtain the redshift. 
In order to simulate the single--binary encounters with ARWV, other parameters are needed, which are described in section \ref{initial_conditions}. Since the distributions of the parameters span a very wide range of values, we also obtained a wide range of characteristic frequencies of the different encounters. On the basis of step-wise refined estimates of these frequencies, we have chosen the simulated duration of the encounters and the temporal resolution used for ARWV.

\subsection{Redshift distribution of the encounters} \label{ne_result}

\begin{figure*}[t]
\includegraphics[width=\textwidth]{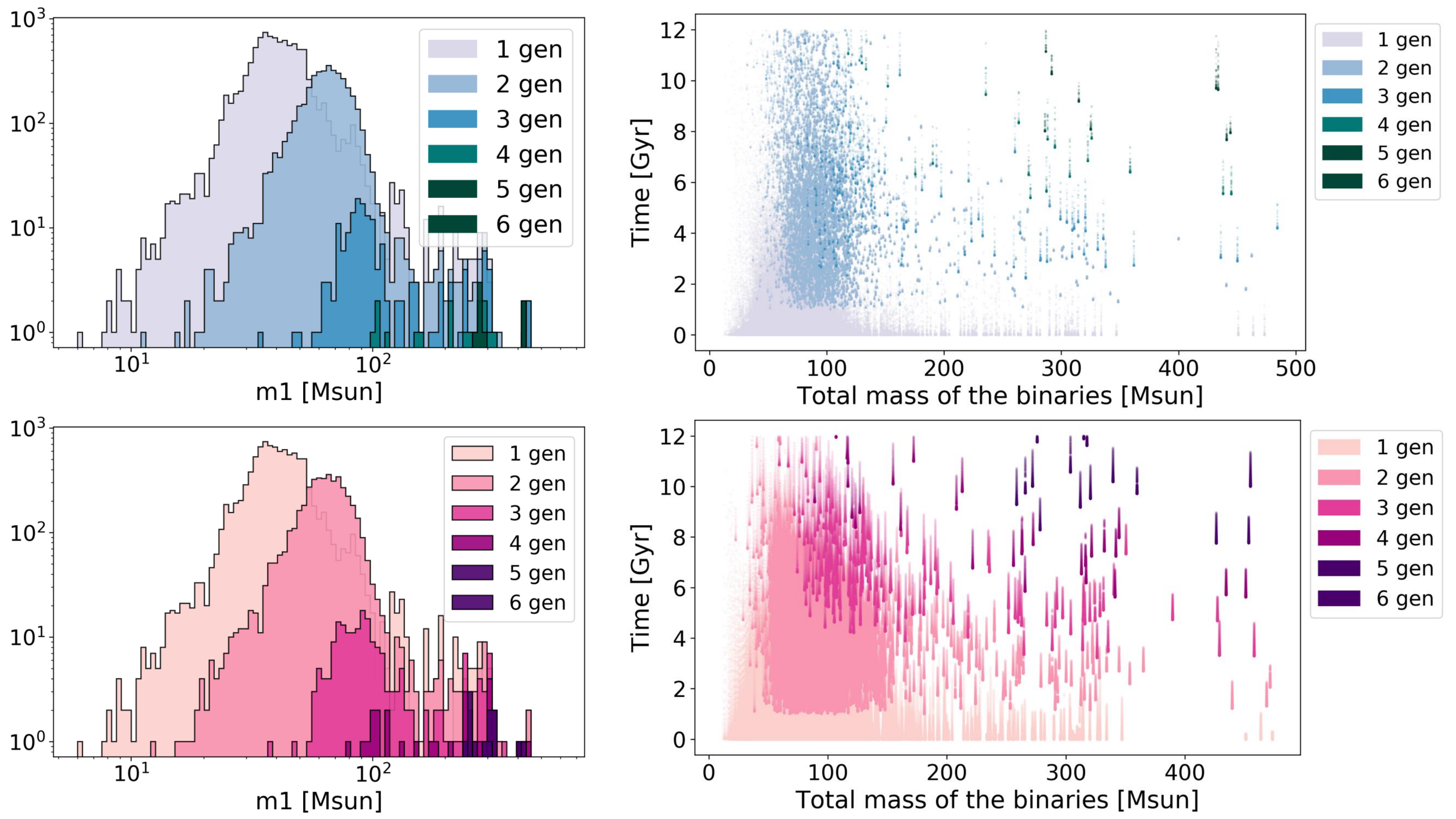}
\caption{Left-hand panels: distribution of the mass values m$_1$ of the binary in the different generations. Masses $m_1$ of the first generation BHs are the most massive component of the binary, that we inject at the beginning, while masses $m_1$ in the following generations are the BHs resulting from the merger of the previous generation binaries when they are retained in the NSC.
The upper panel refers to the scenario where the core radius of the NSC is 0.1\,pc, while the bottom panel refers to the 1\,pc scenario. Right-hand panels: total mass of the binaries involved in an encounter with an object in the environment. The upper panel refers to the case of a core of 0.1\, pc, the lower one to 1\,pc. Unlike the first scenario where the intruders are all BHs, in the case of 1\, pc cores, encounters occur more frequently with stars. In all these simulations there are always $10^4$ first generation BHs.}
\label{binaries}
\centering
\end{figure*}

We explore two scenarios: one in which the size of the core of the NSC has a radius of 0.1\,pc and one in which it is 1\,pc. For simplicity, we consider the density of objects in the core to be $10^6$/pc$^3$; this is a conservative assumption since the true density follows a Bahcall-Wolf distribution \citep{bahcall1976star}. Then, we assume that all the BHs are inside the core. We let binaries evolve for 12\,Gyr according to the procedure described in section \ref{evo}. From these simulations we obtained the following results for the two cases:
\begin{itemize}
\item 0.1\,pc core radius: the binaries make a total of about $2\times 10^6$ encounters with single BHs in the case of $10^5$ starting binaries. 

\item 1\,pc core radius: the binaries make a total of about $3.6\times 10^5$ encounters with single BHs in the case of $10^6$ initial binaries and $3.6 \times 10^4$ encounters in the case of $10^5$ initial binaries.

\end{itemize}
Therefore, starting with the same number of binaries,  we find on average 100 times as many encounters in the scenario of 0.1\,pc core compared to that of 1\,pc. 

The binaries involved in the encounters can still be the original binaries - we call them first generation binaries - but also some of the following generations, composed of a BH resulting from the merger of the previous generation binary and a BH of the cluster, randomly chosen from the list of intruders, i.e. from our distribution of single BHs. Among the various simulations made, we get a maximum of 8 generations. On the right side of Figure \ref{binaries}, each dot of the plot represents an encounter between a binary and an object in the surrounding environment. In particular, the total mass of the binary and the time at which the encounters takes place in the simulation are indicated. The different colors refer to the generation of the binary. First generation binaries are the ones we injected at a time close to the formation of the cluster. From the plots we can see that even these binaries, especially those with lower mass, can last up to the present time (12\,Gyr): in fact, assuming that intruders have all the same mass, and that all binaries start with the same semi-major axis, then, at each encounter, binaries with a smaller total mass would tighten more, because they acquire a larger fraction of energy ($\Delta{}E_b$); therefore, with a smaller semi-major axis, encounters are less frequent. On the other hand, binaries with a greater total mass display a small reduction of the semi-major axis after each encounter, thus making encounters more frequent.

For the following generations, the mass $m_1$ of the binary is the result of the merger of the previous generation binary. The increase of the distribution of $m_1$ as the generations progress is illustrated on the left-hand side of Figure \ref{binaries} for the two scenarios. At the same generation, smaller masses are formed when the core is 1\,pc.


As a result of our simulation, for all the binaries making an encounter with a third BH we have an associated redshift and the semi-major axis of the binary which, due to previous encounters, is shrunk compared to the initial one. 
In order to simulate  the entire single--binary encounter, ARWV will need these parameters as initial conditions, together with those described in Sec. \ref{initial_conditions}, that we can draw randomly from the distributions we defined there.

\begin{figure}[t]
\includegraphics[width=0.98\columnwidth]{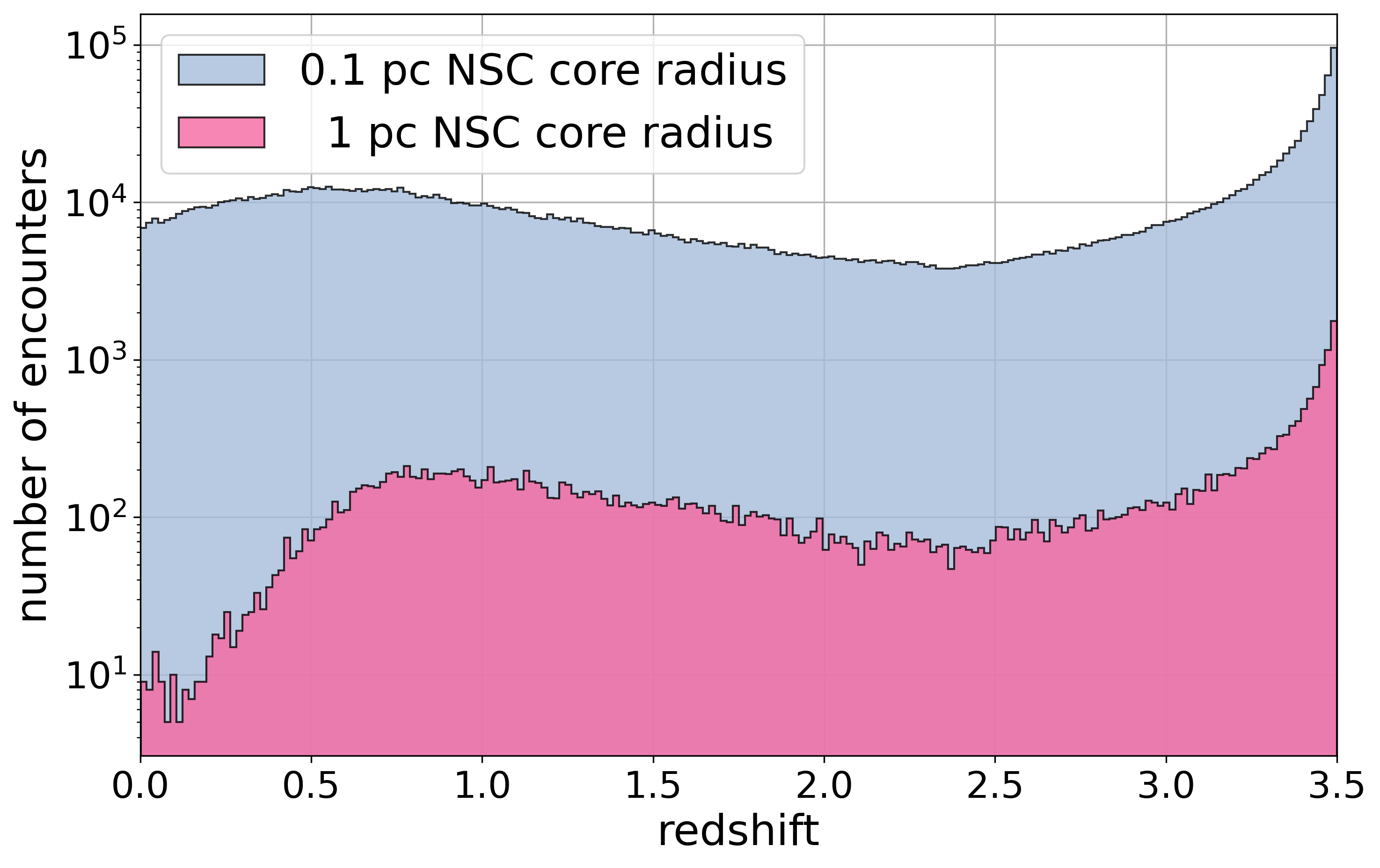}
\caption{Number of single--binary encounters between BHs as a function of redshift in the case of 10$^5$ initial binaries. The pink histogram refers to the scenario of a radius of the NSC core of 1\,pc, the blue one for a core of 0.1\,pc radius. }
\label{redshift_ne}
\centering
\end{figure}

Figure \ref{redshift_ne} shows the histogram of the number of encounters as a function of the redshift in the two scenarios for the case of $10^5$ original binaries. It is evident that, while the trend remains similar in both scenarios, with a 0.1\,pc core there is a greater number of encounters.

\subsection{Gravitational signal} \label{signal}

Simulating a high number of encounters with ARWV to generate spectra of emitted GW signals is computationally very expensive. For this reason, we used ARWV only to produce several examples of GW spectra, and to assess whether analytical approximations could be used to produce these spectra for a large number of signals. We start by presenting the analysis of a selected subset of single--binary encounters whose parameters have values that can generate a GW signal of interest for our interferometers. In particular, among $3.6\times{}10^5$ BBH-BH encounters happening over the course of 12\,Gyr in the 1\,pc core, we have selected four groups of events for a parametric study of the GW signal strength:
\begin{itemize}
\item events within 160\,Mpc. 
\item events in which the mass $m_1$ of the binary is greater than 450 $M_\odot$;
\item  events in which the initial relative velocity $v_0$ between binary and intruder is greater than 190\,km/s;
\item  events in which the semi-major axis of the binary is less than 0.2\,AU.
\end{itemize}

\begin{figure*}[t]
\includegraphics[width=\textwidth]{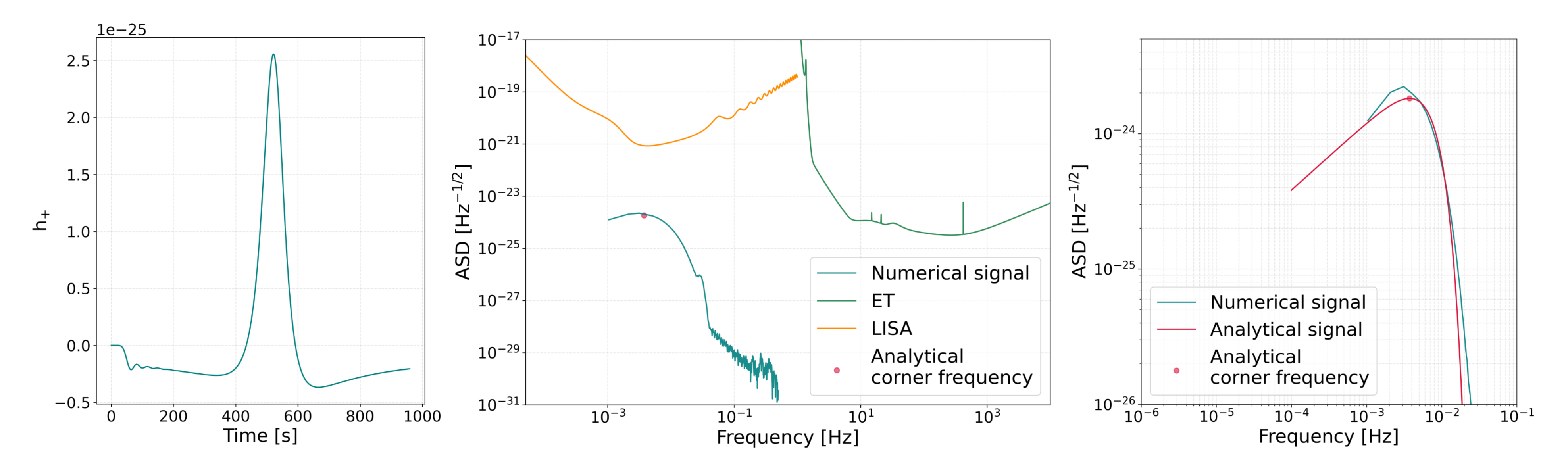}
\caption{The middle plot shows the simulated GW spectrum 2$\sqrt{f}\tilde h(f)$ of the single--binary encounter, whose trajectory is shown in Fig.~\ref{trajectory},
compared to the LISA and ET sensitivity curves. In the left plot, the corresponding GW burst is shown in time domain around the closest distance between two BHs. A comparison of the numerical and simplified analytical models is shown in the right plot.}
\label{signal_re}
\centering
\end{figure*}

Events selected in this way represent only less than 2.5\% of all the encounters. Only the 44\% of this small sample, corresponding to 3980 single--binary encounters, turned out to be purely hyperbolic by looking at the trajectory of the three bodies from the ARWV simulation with a time resolution of 1\,s. Table \ref{tab1} shows in detail the number of encounters selected for each category, along with the absolute number and the relative percentage of hyperbolic events. This is not an indication of a general relationship between hyperbolic and resonant encounters (or exchanges): our set of events is very peculiar and includes cases in which the semi-major axis of the binary is very small, so that the encounter with the intruder will more likely end up in a three-body system that remains bound for a fairly long period of time. 

We calculate the wave amplitude of the signal according to \cite{ferrari2020general}:
\begin{equation}
h^{TT}_{jk}=\dfrac{2G}{c^4 R} \ddot{Q}^{\rm TT}_{jk},  \label{hTT}
\end{equation}
where $j,\,k=1,2,3$, $R$ is the luminosity distance of the source from us, and $\ddot{Q}^{\rm TT}_{jk}$ is the second time derivative of the transverse-traceless part of the quadrupole moment.

We consider the wave propagating along the $z-$direction and, of the full GW tensor, we choose the $h_{11}$ component - the plus polarization -  to calculate the signal spectrum as $2f^{1/2} \mid\tilde{h}_+(f)\mid $ \cite{moore2014gravitational}, where $\tilde{h}_+(f)$ is the Fourier transform of $h_+(t)$.

The middle panel of Figure \ref{signal_re} shows the amplitude spectral density as a function of the frequency of a typical signal calculated in the numerical procedure. We applied a low-pass filter to reduce the numerical error at high frequency coming from the ARWV simulations.
In the example shown in Figure \ref{signal_re}, the threshold frequency of the low-pass filter is $3\times 10^{-2}$\,Hz. 

The trajectories of the three bodies that generate the signal in Figure \ref{signal_re} are plotted with a time resolution of 1\,s in Figure \ref{trajectory}. The masses of the binary components are 94.5\,M$_{\odot}$ and 10.9\,M$_{\odot}$, while the mass of the intruder is 7.1\,M$_{\odot}$. The two-body hyperbolic encounter occurs between the intruder and the most massive component of the binary. They reach a minimum distance of 0.002\,AU with  relative velocity of $8\times 10^3$\,km/s. At that moment, the binary separation is 0.128\,AU. The luminosity distance at which the encounter happens is 2.1\,Gpc. 

We compared the numerical simulation of the GW burst with a simple analytical model, i.e., a Gaussian-shaped burst in the time domain, due to a two-body hyperbolic encounters \citep{8capozziello2008gravitational,99garcia2018gravitational}. Its maximum amplitude, at the moment of closest approach, is $h_+$, which we get from equation (\ref{hTT}) at the characteristic frequency. The Fourier transform $\tilde h(f)$ of this signal is again a Gaussian, and we compute the amplitude spectral density 2$\sqrt{f}\,{}\tilde h(f)$ to compare it with the detector sensitivities. This simplified model matches well our numerical signals around peak amplitude when the single--binary encounter is purely hyperbolic between two of the three bodies, with the third body being far enough so that its effect on the trajectory of the encounter is negligible. This is the case of the signal in Figure \ref{signal_re}: in the left panel of Fig \ref{signal_re} the filtered numerical signal in the time domain is shown. 

With the red dot in Figure \ref{signal_re}, we identify the amplitude spectral density at the corner frequency defined in Eq.~\ref{f}. To avoid the numerical simulation of all encounters, we use this as estimate of the peak amplitude of the signals. Comparing with numerical estimates for several signals, we find that the analytical estimate is similar or smaller by a modest factor compared to the numerical peak amplitudes.
The underestimated analytical signals are due to the fact that our model is a Gaussian approximation of the time domain signal using the minimum distance between bodies in the ARWV simulations. Signals from encounters with spinning black holes can have larger peak amplitudes \citep{de2014gravitational}. For a reference set of signals, we find that peak amplitudes calculated with the numerical simulation are larger by up to a factor 5 for almost all of the signals.

\begin{figure*}[t]
\centering
\includegraphics[width=\textwidth]{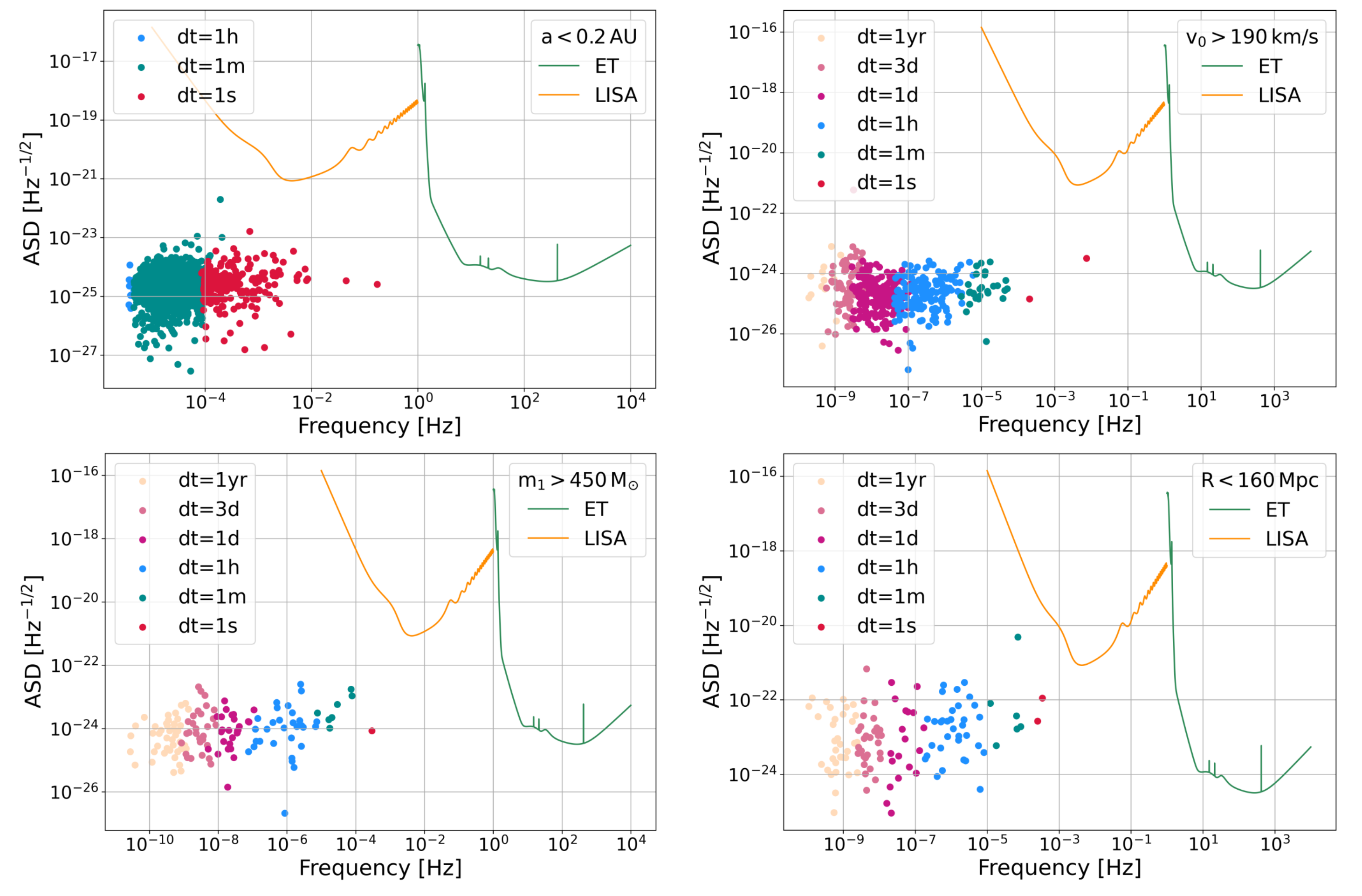}
\caption{Characteristic frequency and amplitude spectral density of the GW emitted during single--binary encounters with respect to the sensitivity curves of LISA and ET. Each dot represents an encounter in which the intruder makes a flyby around one object of the binary. The coordinates of the dots are extracted from the analytical signal at the characteristic frequency defined in Eq \ref{f}. The different colors refer to the different time resolutions we used to simulate them with ARWV. We selected four samples of events according to specific value of the parameters in the scenario of 1\,pc core radius: in the top right-hand panel, the relative velocity at infinity between the binary and the intruder is higher than 190\,km/s. Top left-hand panel:  the semi-major axis of the binary is smaller than 0.2\,AU. Lower left-hand panel: the most massive component of the binary exceeds 450 M$_{\odot}$. Lower right-hand panel: the distance of the encounter from us is less than 160\,Mpc.}
\label{dots_separate}
\end{figure*}

We plot these signals in Figure \ref{dots_separate}. Since the analytical approximation is valid for two-body encounters, we took into account only hyperbolic events for which, at the moment of maximum approach, the distance between the two masses of the binary is at least triple the distance between the intruder and the mass around which it flies by. This allows us to consider only scenarios that are more similar to an encounter between two bodies, making it meaningful to compare the numerical signal to the analytical one.
The signals in Fig. \ref{dots_separate} are in total - for the four categories - 1863 out of the set of 3980 hyperbolic encounters. The different colors refer to the different time resolutions that we used to simulate the encounters. They are categorized according to the values of the parameters that we selected. Moving clockwise from the top left panel, there are the encounters during which the semi-major axis of the binary is smaller than 0.2\,AU, those in which at infinity the relative velocity between the binary and the intruder is higher than 190\,km/s, those that are closest to us, i.e. within a luminosity distance of 160\,Mpc, and finally, the ones for which the mass $m_1$ of the binary exceeds 450 M$_{\odot}$ (this does not imply that the closest approach of the intruder happens with this object).

The highest frequency signals are those in which the semi-major axis of the binary is smaller, even though none of the signals fall inside the sensitivity curves of the two interferometers.
This is also the case in which we neglect a larger fraction of encounters because they are resonant: in fact, when the semi-major axis of the binary is smaller, it is more likely that an unstable bound system of three bodies is formed, ending up with the ejection of one of the three. This situation is promising in view of finding signals in the ET band, when we will also analyze the resonant signals.


\begin{table}
  \centering
  \begin{tabular}{P{2.5cm} P{1.3cm} P{1.3cm} P{1.3cm}}
  \hline
  \hline
    \multicolumn{4}{c}{1\,pc NSC core, 3.6$\times$10$^5$ BBH-BH encounters } \\
    \hline
    Category & Total & Flyby & $\%$\\
    \hline
    a$_{binary} <$ 0.2\,AU & 7561 & 2750 & 36.4$\,\%$ \\
    v$_0 >$ 190\, km/s & 861 & 760 & 88.3$\,\%$ \\
    m$_1 >$ 450\,M$_{\odot}$ & 287 & 245 & 85.4$\,\%$ \\
    R $<$ 160\,Mpc & 284 & 225 & 79.2$\,\%$ \\
    \hline
  \end{tabular}
  \caption{Values of the parameters that we used to select the events. For the scenario of 1\,pc NSC core radii, starting with 10$^6$ first generation binaries we get a total of 3.6$\times$10$^5$ BBH-BH encounters; of these we selected a subset according to specific values of the parameters, reported in the first column. The total number of events per category is shown in the second column. The number of encounters in which the intruder flies by the binary are in the third column while their percentage is reported in the fourth column. 
  }\label{tab1}
\end{table}

\subsection{Rate of encounters} \label{rate}

The number of single--binary encounters per year is computed as:
\begin{equation}
R(z)=  \int_{z_{\text{min}}}^{z_{\text{max}}} R_m(z) \dfrac{dV(z)}{\text{d}z} \dfrac{1}{1+z}\,dz 
\end{equation}
where $R_m(z)=dN/dt/dV$ is the
source-frame rate density at redshift $z$, $dV/\text{d}z$ is the differential comoving volume shell and $(1+z) ^{-1} $ accounts for the time dilation due to cosmic expansion between the source and the observer frames.

\begin{figure}[t]
\includegraphics[width=0.98\columnwidth]{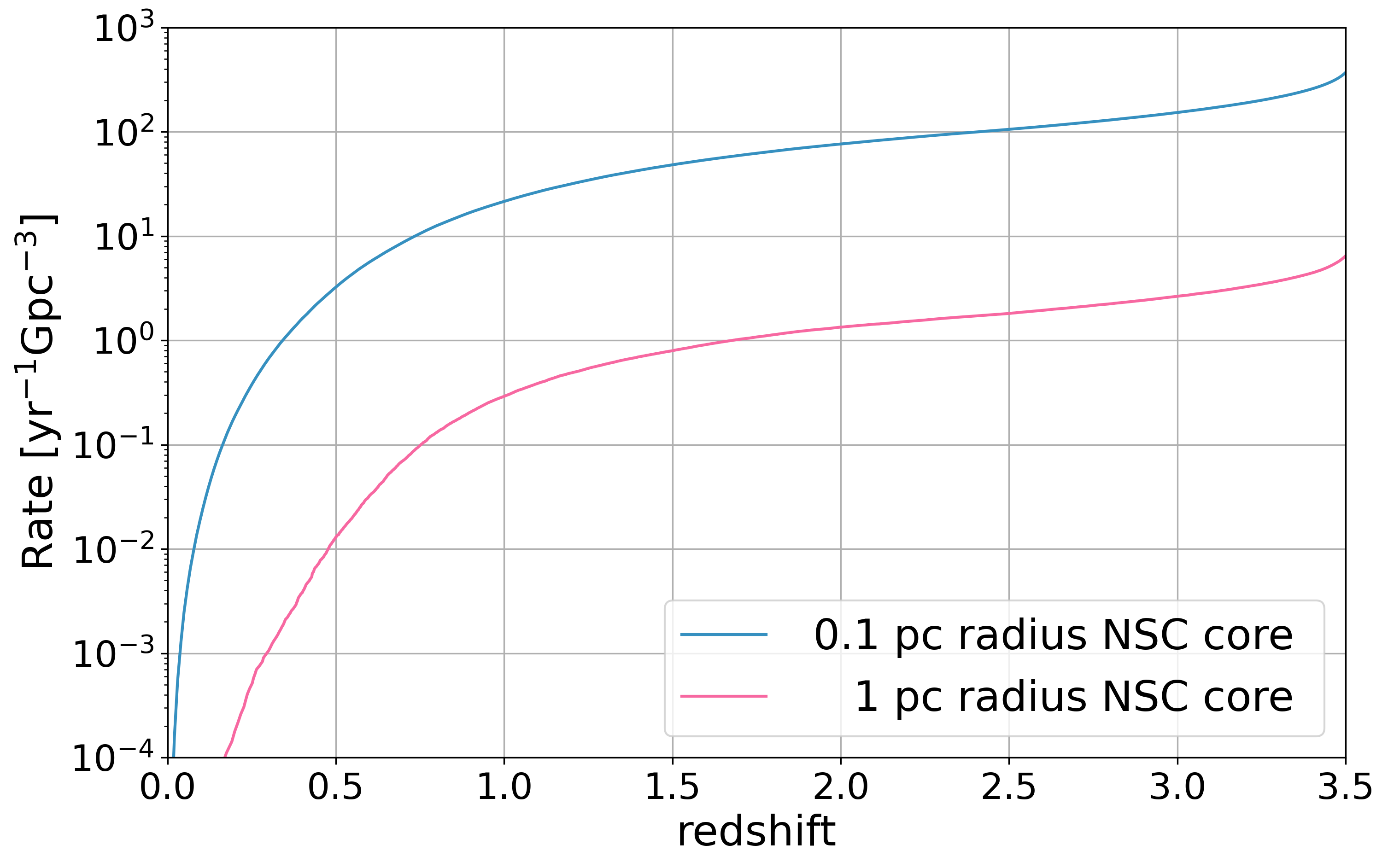}
\caption{Cumulative rate of the expected number of single--binary encounters between BHs per year. The blue curve refers to the scenario of a radius of the NSC core of 0.1\,pc while the pink curve for a core of 1\,pc radius. }\label{cumulative_rate}
\centering
\end{figure}

In Figure \ref{cumulative_rate} we show the trend of the cumulative rate in the two scenarios that we have considered, which can be seen as upper and lower limit of the number of single--binary encounters that we expect to happen in NSCs up to a certain redshift. The rate up to $z\,=\,3.5$ then is within the range $[0.006 - 0.345]\,$yr$^{-1}\,$Gpc$^{-3}$.

\begin{figure}[t]
\includegraphics[width=0.98\columnwidth]{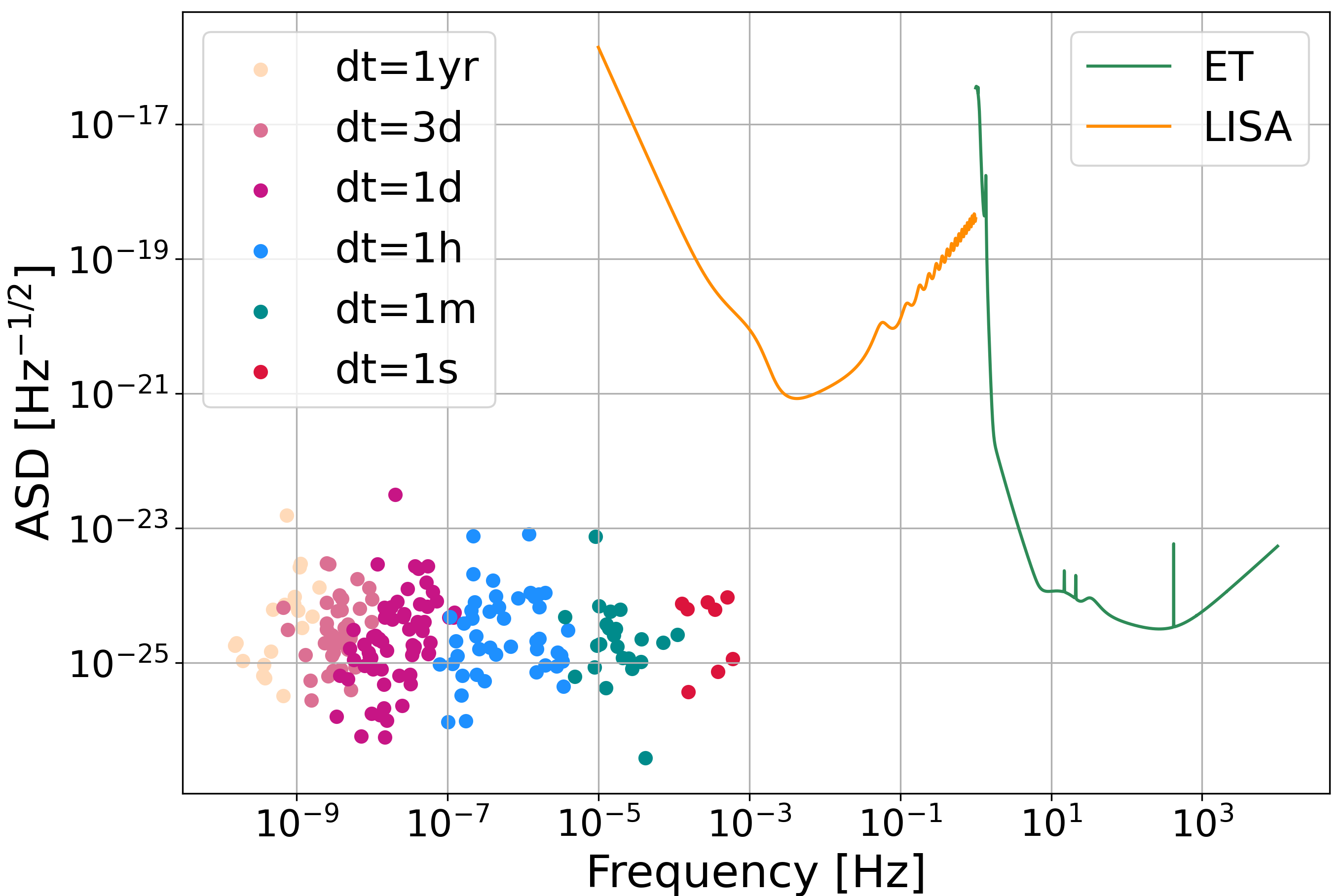}
\caption{Same as Fig.~\ref{dots_separate}. Here the events are extracted according to the distribution in Fig.~\ref{cumulative_rate} for the scenario of 0.1\,pc core radius. We simulate a total of 542 encounters, i.e. the number of encounters that we expect to happen in one year up to redshift 3.5. The 289 events plotted are the purely hyperbolic for which at the time of the encounter the distance between the two components of the binary is at least triple the distance between the intruder and the body it passes closest. }\label{sample_1yr}
\centering
\end{figure}

From Figure \ref{cumulative_rate} we observe, for the case of 0.1\,pc NSC radius, a cumulative rate of about 500 events per year. Precisely, we extrapolate 542 events following the distribution in Figure \ref{redshift_ne} and we simulate them with ARWV in order to measure their frequencies and spectrum.
Of these 542 encounters, 365 are hyperbolic. 
For those in which the minimum distance between the intruder and the closest body of the binary is less than one third the distance between the two components of the binary at that moment, the amplitude spectral density is plotted in Figure \ref{sample_1yr} as a function of their characteristic frequencies.

\begin{figure}[t]
\includegraphics[width=1\columnwidth]{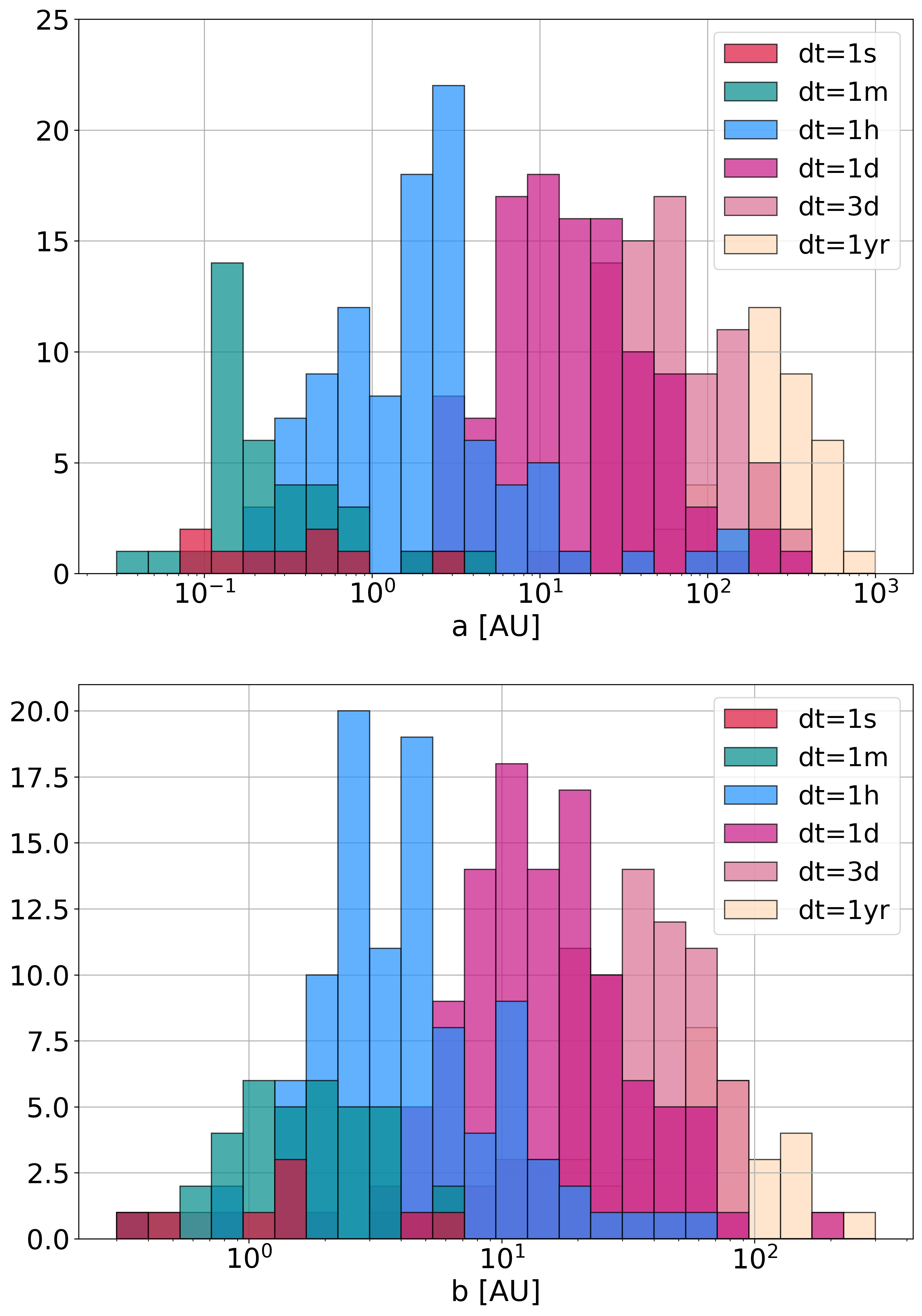}
\caption{Distribution of the initial values of the impact parameters with which the intruder approaches from an infinite distance (plot below) and of the semi-major axes of the binaries before the encounter (plot above). These are the 365 hyperbolic events that we used to obtain the plot in Figure \ref{sample_1yr}. From Fig \ref{sample_1yr} it can be seen how events with higher frequency, which we simulated with a lower temporal resolution, are those for which the initial semi-major axis and the impact parameter of the encounter are lower.}\label{hist}
\centering
\end{figure}

By comparing the signals obtained (Fig. \ref{sample_1yr}) with the values of the initial parameters that describe the encounter (Fig. \ref{hist}), we notice a strong dependence of the characteristic frequency of the encounter from the initial values of the semi-major axis and from the impact parameter of the binary; from Fig. \ref{hist} it is evident that encounters with small semi-major axis and small value of the impact parameter, which are those we simulate with a lower time resolution, have a higher frequency. We have not noticed the same dependence by looking at other types of parameters such as the masses, the mass ratio, the initial velocity of the intruder, or the generation number to which the binary belongs.

\section{Conclusions}\label{section6}
We presented a calculation of the rate of encounters between three black holes (BHs) in nuclear star clusters (NSCs) as a function of redshift. In particular, we focused on single--binary encounters which, in dense stellar environments, we assumed to occur more frequently than encounters between two or three unbound BH \citep{portegies2000,banerjee2010,tanikawa2013,oleary, mapellietal2013, ziosietal,rodriguez2015,rodriguez2016,rodriguez2019, mapelli2016,askar2017,seddaetal,samsing2018,samsing2018b,fragione2018,fragione2019,fragione2022,zevin2019,zevin2021,kremer2019, Sedda_2020, mapellietal2021, rastello,banerjee2021,rizzuto2022,kamlah2022}.
Moreover, we selected the hyperbolic encounters, since the resonant encounters require more careful numerical analyses to calculate the associated GW signals and their impact on the rate. 

The probability of BBH-BH encounters depends strongly on the core size of the NSC. We found that for a core size of 1\,pc, we can expect about 10 BBH-BH encounters per year up to redshift $z=3.5$, and for a core size of 0.1\,pc, which means increased BH density, we find that about 500 BBH-BH encounters occur. 

The hyperbolic encounters were selected using the N-body code ARWV, from which we obtain the positions and the velocities of the three bodies at each time step to provide an analytical estimate of the GW signal amplitude. The simplified analytical estimates were compared with the full numerical results for several encounters to confirm a match between the two. We found that the vast majority of the encounters have their peak GW emission below the LISA sensitivity band. Several signals appear in the LISA band, while no signal was found in the observation band of present and future terrestrial GW detectors. We found the highest-frequency peak emission close to 0.1\,Hz. The signals in the LISA band have too low amplitude to be detectable. This leads us to the conclusion that GW signals from hyperbolic BBH-BH encounters in NSCs will likely remain undetected in the foreseeable future. 

Our initial work leaves possibilities for follow-up studies. First, resonant encounters are very promising candidates for detectable GW signals, but they require a more demanding numerical analysis, which is why we omitted them in this first study. Also, resonant encounters will give rise to complex GW signals, and the question of how one would detect them needs to be addressed carefully as well. Second, our study focused on NSCs, which show stronger mass segregation and high density in their cores making BBH-BH encounters more likely. However, whether most BBH-BH encounters happen inside NSCs should be tested more carefully since other types of star clusters are more numerous than NSCs. Furthermore, while building our distributions of BHs (both binary and single) we considered masses coming from the astrophysical evolution of stars in a dynamical environment \citep{dicarloetal}, neglecting the possible presence of more massive primordial BHs \citep{greene2012low, clesse2017clustering}. 
We also neglected the possible interactions between the binaries with the supermassive BH that could be at the center of galaxies \citep{volonteri2010formation}, as well as with possible intermediate mass BHs that are believed to populate the cores of NSCs \citep{miller2012upper,mckernan2012intermediate}.
Finally, we made the simplistic assumption of considering the properties of the cluster to be stationary during the entire simulation of 12\,Gyr. For NSCs this choice is conservative: during its life, the cluster could increase its mass over time due to new star formation \citep{mapelli2012situ} and by accreating globular clusters (\cite{capuzzo2008self}, \cite{antonini2012dissipationless}).

\section{Acknowledgements}
The authors acknowledge use of the CalTech LIGO Cluster hosted by the California Institute of Technology, on which some of the numerical computations for this paper took place. MM acknowledges financial support from the European Research Council for the ERC Consolidator grant DEMOBLACK, under contract no. 770017. MD acknowledges financial support from Cariparo foundation under grant 55440.

\bibliographystyle{apsrev4-2} 
\bibliography{references} 

\end{document}